\documentclass[preprint2]{aastex}

\def\cm{cm$^{{\rm -1}}\,$}
\def\TR#1#2#3#4#5#6#7#8{#1 $^{\rm #2}$#3$_{\rm #4}$ -- #5
$^{\rm #6}$#7$_{\rm #8}$}

\begin{document}

%
%
\title{Interferometric measurement of resonance transition wavelengths
in \ion{C}{4}, \ion{Si}{4}, \ion{Al}{3}, \ion{Al}{2}, and \ion{Si}{2}}

%
%
\author{Ulf Griesmann\altaffilmark{1,2} and Rainer Kling\altaffilmark{1}}
\altaffiltext{1}{National Institute of Standards and Technology (NIST),
Gaithersburg, Maryland 20899, U.S.A.}
\altaffiltext{2}{Harvard--Smithsonian Center for Astrophysics,
Cambridge, Massachusetts 02138, U.S.A.}

\email{ulf.griesmann@nist.gov, rainer.kling@nist.gov}

%
%
\shortauthors{Griesmann and Kling} 

\shorttitle{\ion{C}{4}, \ion{Si}{4}, \ion{Al}{3}, \ion{Al}{2}, and
\ion{Si}{2} resonance transition wavelengths}

%
%
\begin{abstract}
We have made the first interferomeric measurements of the wavelengths
of the important ultraviolet diagnostic lines in the spectra
\ion{C}{4} near 155\,nm and \ion{Si}{4} near 139\,nm with a vacuum
ultraviolet Fourier transform spectrometer and high-current discharge
sources. The wavelength uncertainties were reduced by one order of
magnitude for the \ion{C}{4} lines and by two orders of magnitude for
the \ion{Si}{4} lines. Our measurements also provide accurate
wavelengths for resonance transitions in \ion{Al}{3}, \ion{Al}{2}, and
\ion{Si}{2}.
\vspace{5mm}
\end{abstract}

\newpage

%
%
\noindent
The resonance transitions \TR{2s}{2}{S}{1/2}{2p}{2}{P}{1/2,3/2} near
155\,nm in \ion{C}{4} and \TR{3s}{2}{S}{1/2}{3p}{2}{P}{1/2,3/2} near
139\,nm in \ion{Si}{4} are among the most important transitions for
optical plasma diagnostics of hot plasmas in the interstellar medium,
the intergalactic medium and stellar atmospheres. During the past
decade, the increasing light gathering power of ground and space based
telescopes has made it feasible to observe these lines with high
resolution spectrometers. For example, the Goddard High Resolution
Spectrograph (GHRS) and the Space Telescope Imaging Spectrograph
(STIS) on the Hubble Space Telescope have modes of operation with
resolving powers of up to 100000. For a reliable analysis of high
resolution spectra from these instruments, transition wavelengths must
be known with an accuracy of at least 1 part in 10$^{\rm 6}$.  We have
therefore undertaken new measurements of the \ion{C}{4} and
\ion{Si}{4} resonance transition wavelengths. Our results confirm the
observation of \citet{smith_98} that the accuracy of available
wavelength data for doubly- and triply-ionized atoms is often very
inadequate.

The strongest motivation, at present, for a new measurement of the
\ion{C}{4} and \ion{Si}{4} resonance lines came from the recent
interest in determining bounds for temporal and/or spatial variations
of the fine structure constant. This may be achieved through a
measurement of the splitting of spectral lines in fine structure
multiplets through spectroscopic observations of distant gas clouds
seen against a background quasar.  In a series of articles
\citep{webb_99,dzuba_99a,dzuba_99b}, a method was suggested that
considerably increases the sensitivity of optical measurements to
variations of the fine structure constant.  A recent analysis of
quasar absorption spectra \citep{webb_99}, based on accurate
wavelengths for {\em ultraviolet} transitions in \ion{Mg}{2}
\citep{pickering_98} and \ion{Fe}{2} \citep{nave_91}, found weak
statistical evidence for a smaller fine structure constant in systems
with cosmological redshifts $z > 1$. With this paper we intend to
contribute accurate measurements of wavelengths for resonance
transitions of several important ions in the {\em vacuum ultraviolet}
to enable the analysis of quasar absorption spectra with much larger
cosmological redshifts.

\begin{figure}
\plotone{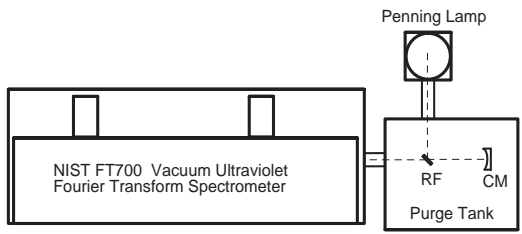}
\caption{Experimental setup. Light from a high-current Penning
discharge is reflected onto a concave mirror (CM) by a
metal-dielectric reflection filter (RF), which is located below the
optical axis of the spectrometer. The concave mirror images the
Penning discharge near the entrance aperture of the Fourier transform
spectrometer.}
\label{fig_setup}
\end{figure}

Figure \ref{fig_setup} is a block diagram of the experimental setup.
The spectra were measured with the FT700 vacuum ultraviolet Fourier
transform spectrometer (FTS) at the National Institute of Standards
and Technology (NIST) \citep{griesmann_99}. Photomultipliers with
cesium-iodine photocathodes limited the optical bandpass of the
spectrometer to wavelengths below $\sim$190\,nm which was essential to
achieve sufficient signal-to-noise ratios in the vacuum ultraviolet. A
number of scans were co-added for total observation times of 20\,min
to 30\,min per interferogram.  A high-current Penning discharge lamp
\citep{heise_94} with silicon carbide (SiC) cathodes and Ne as carrier
gas was used to excite Si and C spectra.  The SiC cathodes of our
Penning lamp contained Al as an impurity in sufficient quantity to
allow us to measure the wavelengths of the \ion{Al}{3} and \ion{Al}{2}
resonance transitions.  High discharge currents up to 2\,A and carrier
gas pressures as low as 0.2\,Pa were needed to excite spectra of
triply ionized atoms in the Penning discharge.  The radiation emitted
from the center of the discharge was imaged near the entrance aperture
of the Fourier transform spectrometer with a concave mirror (CM in
Figure \ref{fig_setup}) and a metal-dielectric reflectance filter (RF
in Figure \ref{fig_setup}). To avoid systematic uncertainties
resulting from inhomogeneous illumination of the entrance aperture,
the location of the discharge image was displaced from the entrance
aperture by several cm to achieve a degree of spatial averaging in the
plane of the entrance aperture. The aperture ratio of the imaging
system was approximately matched to that of the spectrometer.  All
components of the imaging system were enclosed in a sealed chamber
that was continuously purged with Ar gas to replace oxygen in the
optical path. The metal-dielectric reflectance filter had a peak
transmission of 60\% at 160\,nm, near the \ion{C}{4} resonance lines,
and a bandpass (FWHM) of approximately 30\,nm. The purpose of this
filter was to reduce the intensity of several very strong \ion{Si}{2}
lines around 180\,nm and thus to increase the signal-to-noise ratio in
the spectra. Without this filter the weak \ion{C}{4} resonance lines
were not observed.  The Doppler width of spectral lines emitted by the
Penning lamp under high current conditions was approximately 1\,\cm\,
and thus a spectral resolution of $\sim$0.2\,\cm was sufficient to
resolve them. The interferograms were phase corrected, transformed and
analyzed with the FTS data analysis program ``xgremlin''
\citep{nave_97}.  

It is one of the greatest strengths of the FTS method that the large
optical bandpass allows the wavenumber calibration of VUV spectra with
accurate wavenumber standards at much smaller wavenumbers. Wavenumber
calibration is global, not local as with grating spectrometers. This
leads to short calibration chains and therefore smaller statistical
and systematic uncertainties.  The wavenumber calibration of our
spectra is based on a selection of the \ion{Ar}{2} wavelength standard
lines near 400\,nm which were measured by \citet{whaling_95}. A broad
band spectrum of the Penning lamp operating with Ar as a carrier gas
and at a relatively low current of 0.5\,A was measured. We selected 24
standard lines between 430\,nm and 520\,nm with signal-to-noise ratios
between 200 and 1000 to calibrate the broad band spectrum. All lines
had upper levels of low excitation and were free of self absorption.
This resulted in very accurate wavenumbers for two \ion{Si}{2} lines
at 181.693\,nm and 180.801\,nm (see Table
\ref{tab_wavenumbers}). These lines, which have a signal-to-noise
ratio of at least 500 in both low and high current spectra, were then
used to calibrate the wavenumber scales of the spectra from the
high-current measurements in the vacuum ultraviolet.

\begin{figure}
\plotone{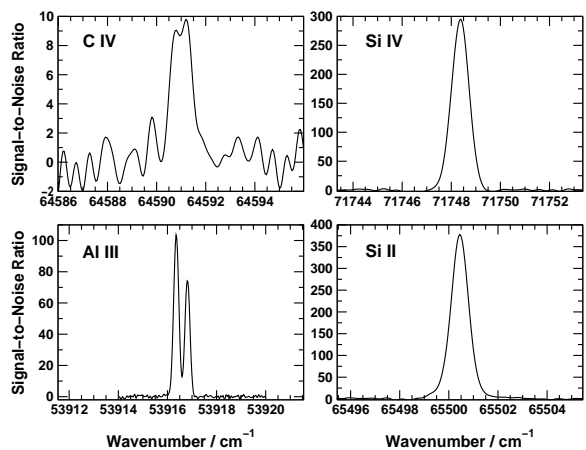}
\caption{Spectral lines of resonance transitions in \ion{C}{4},
\ion{Si}{4}, \ion{Al}{3}, and \ion{Si}{2}. The spectra are normalized
such that the {\it rms} noise level is equal to unity. The abscissae
of the plots show a range of 10\,\cm. The \ion{Al}{3} lines were measured
with a hollow-cathode lamp.}
\label{fig_spectra}
\end{figure}

Table \ref{tab_wavenumbers} contains our results for resonance
transition wavenumbers in \ion{C}{4}, \ion{Si}{4}, \ion{Al}{3},
\ion{Al}{2}, and \ion{Si}{2}. The wavenumbers shown are the combined
results of several measurements that were made for different Penning
discharge conditions. Five interferograms were analyzed for \ion{C}{4}
and eight for \ion{Si}{4}.  The table includes values from widely used
critical compilations of optimized energy levels for comparison
\citep{martin_79,martin_83,moore_93}.  For the \ion{C}{4} lines we
have given the uncertainty of the best previous measurement by
\citet{bockasten_63}. Uncertainties of the previous wavenumber data
for the other ions can be found in the compilation by
\citet{kaufman_74}. Figure \ref{fig_spectra} shows the signal-to-noise
ratio that was obtained for several of the transitions.  All but the
\ion{C}{4} lines have very good signal-to-noise ratios of at least
100.  The weakness of the \ion{C}{4} lines reflects the large amount
of energy that is required to create C$^{\rm 3+}$ in collisions. Using
the ionization potentials for multiply charged ions published in
\citet{moore_49} we estimate that $\sim$83.5\,eV must be expended to
remove three electrons from a C atom whereas only $\sim$57.9\,eV are
needed to remove the same number of electrons from a Si atom.  Even
without detailed consideration of the cross sections for the
ionization processes, a higher density of Si$^{\rm 3+}$ than C$^{\rm
3+}$ is thus expected in the Penning discharge.

Spectral line positions were determined by fitting Voigt functions to
the spectral lines. The values in Table \ref{tab_wavenumbers} are the
means of up to eight separate measurements. Uncertainties in Table
\ref{tab_wavenumbers} are reported at a level of one standard
deviation. Statistical uncertainties of the line positions were
calculated as described by \citet{brault_88}. The total statistical
uncertainty of a spectral line position combines the uncertainty in
the wavenumbers of the \ion{Si}{2} lines that were used to calibrate
the spectra in the vacuum ultraviolet and the uncertainties of the
wavenumber scale calibrations. In all cases the statistical line
position uncertainty of the vacuum ultraviolet lines was the dominant
uncertainty component. At first glance, our results in Table
\ref{tab_wavenumbers} are suggestive of systematic errors in our
measurements because they are all somewhat smaller than previous data.
The two systematic errors that could lead to significant uncertainties
are pressure shifts in the discharge and illumination
shifts. \citet{learner_88} have investigated both effects in FTS
measurements with a hollow cathode lamp and found that they never
exceed a few times 10$^{\rm -3}$\cm. Pressure shifts in the Penning
discharge lamp are negligible because the pressure in the lamp is at
least two orders of magnitude lower than in a hollow cathode
lamp. Finally, the Zeeman splitting of the spectral lines emitted in
the magnetic field of the Penning lamp (0.09\,T) is only
$\sim$0.1\,\cm - about an order of magnitude smaller than the Doppler
width of the lines. The Zeeman splitting is symmetric and does not
shift the line positions.

The Al and Si lines included in Table \ref{tab_wavenumbers} also allow
us to make a strong presumptive case for the absence of significant
systematic uncertainties. If the differences in the last column of
Table \ref{tab_wavenumbers} were the result of a calibration error,
these differences would be proportional to the wavenumber which, as
the numbers clearly show, is not the case. Also, the fact that the
\ion{Si}{2} lines and the \ion{Al}{2} line are in very good agreement
with earlier, relatively accurate, measurements precludes the
possibility of a scale error due to calibration problems. As an
additional test we have measured the \ion{Al}{3} resonance lines with
a high-current hollow-cathode lamp. In this measurement the
\ion{Al}{3} lines were measured together with Ar standard lines. The
results agreed well with those that were obtained with the Penning
lamp.  The discharge temperature in the hollow-cathode lamp is much
lower than in the Penning lamp which allowed us to resolve the
hyperfine structure of the \ion{Al}{3} resonance lines. Natural Al
consists of a single stable isotope $^{\rm 27}$Al which has a nuclear
spin of $I=5/2$. The ground level of Al$^{\rm 2+}$ is split into two
levels with total angular momenta $F=2$ and $F=3$. Table
\ref{tab_wavenumbers} contains wavenumbers for both hyperfine
structure components of the \ion{Al}{3} resonance lines and their
centers of gravity (the wavenumbers were weighted with the integral of
the spectral lines) for comparison with measurements in which the
hyperfine structure remained unresolved.

We conclude that the reason for the significant discrepancies between
our results and previous measurements are the improved experimental
methods developed in the intervening years.  For example, the last
laboratory measurement of the wavelengths of the \ion{C}{4} resonance
lines was made nearly four decades ago by
\citet{bockasten_63}. \citet{tunklev_97} who recently published a new
analysis of the \ion{C}{4} spectrum adopted the wavelengths of
\citet{bockasten_63} for the resonance transitions. The only recent
measurement of the \ion{C}{4} resonance lines by \citet{rottman_90}
was an observation of the lines in the solar photosphere which may be
more susceptible to systematic errors than a laboratory
measurement. In general, the older data appear to have considerable
systematic wavelength uncertainties and those uncertainties seem to
increase with increasing ionization stage. We believe that our new
data make a strong case for further interferometric laboratory
measurements of vacuum ultraviolet spectral lines that are important
for astrophysical plasma diagnostics and for the need to extend the
FTS method to even shorter wavelengths.

%
%
\acknowledgments We gratefully acknowledge financial support by the
National Aeronautic and Space Administration (NASA) through the UVGA
interagency agreement W--19301. We are grateful to Prof.\ Victor
V. Flambaum and Dr.\ John K. Webb, University of New South Wales,
Australia, who encouraged us to undertake the measurements described
in this paper.  U. Griesmann was supported through NIST contract
number 43SBNB960002 with Harvard College Observatory.

%
%


%
%

\begin{deluxetable}{llllll}
\tablecolumns{6}
\tablewidth{0pc}
\tablecaption{Resonance transitions\label{tab_wavenumbers}}
\tablehead{
\colhead{Spectrum} & \colhead{Transition} & \colhead{Wavenumber} &
\colhead{Wavelength} & \colhead{Prev. Value} & \colhead{$\sigma - \sigma_{\rm prev}$} \\
\colhead{}         & \colhead{}           & \colhead{\cm}        &
\colhead{nm}         & \colhead{\cm}          & \colhead{\cm} }
\startdata
\ion{C}{4}  & \TR{2s}{2}{S}{1/2}{2p}{2}{P}{1/2} & 64483.65(9)   & 155.0781(2)    & 64484.0(4)$^a$ & -0.35 \\
            & \TR{2s}{2}{S}{1/2}{2p}{2}{P}{3/2} & 64590.99(4)   & 154.8204(1)    & 64591.7(4)$^a$ & -0.71 \\
\\
\ion{Si}{4} & \TR{3s}{2}{S}{1/2}{3p}{2}{P}{1/2} & 71287.376(2)  & 140.277291(4)  & 71287.5(3)$^b$ & -0.16 \\
            & \TR{3s}{2}{S}{1/2}{3p}{2}{P}{3/2} & 71748.355(2)  & 139.376018(4)  & 71748.6(3)$^b$ & -0.29 \\
\\
\ion{Al}{3} & \TR{3s}{2}{S}{1/2}{3p}{2}{P}{1/2} & 53682.6692(12)& 186.279858(4)  &              & \\
            &                                   & 53683.1953(15)& 186.278033(5)  &              & \\
            &                                   & {\it 53682.884(2)} & {\it 186.279113(7)} & 53682.93(6)$^c$ & -0.046 \\
            & \TR{3s}{2}{S}{1/2}{3p}{2}{P}{3/2} & 53916.3574(6) & 185.472470(2)  &              & \\
            &                                   & 53916.8149(8) & 185.470896(3)  &              & \\
            &                                   & {\it 53916.544(1)} & {\it 185.471829(3)} & 53916.60(6)$^c$ & -0.056 \\
\\
\ion{Al}{2} & \TR{3s$^2$}{1}{S}{0}{3s\,3p}{1}{P}{1} & 59851.976(4)& 167.07886(1) & 59852.02(4)$^c$ & -0.044  \\
\\
\ion{Si}{2} & \TR{3p}{2}{P}{1/2}{3p}{2}{D}{3/2} & 55309.3404(4) & 180.801288(1)  & 55309.35(3)$^b$ & -0.0096 \\
            & \TR{3p}{2}{P}{1/2}{4s}{2}{S}{1/2} & 65500.4538(7) & 152.670698(2)  & 65500.47(4)$^b$ & -0.0138 \\
\enddata

\tablecomments{The wavenumbers of the \ion{Al}{3} resonance lines were
measured with a hollow-cathode lamp; the hyperfine structure of the
lines is discussed in the paper. In the table, the center of gravity
of the hyperfine structure components is given in italics. This value
should be used for comparison with previous measurements which did not
resolve the hyperfine structure. Previous experimental data referred
to in column five are taken from the following compilations:
$^a$\,\citet{moore_93}, $^b$\,\citet{martin_83},
$^c$\,\citet{martin_79}.}
\end{deluxetable}

%
%

\end{document}